\documentclass[9pt,conference,a4paper]{IEEEtran}

\IEEEoverridecommandlockouts
\usepackage{ifpdf}
\ifpdf
  \usepackage[pdftex]{hyperref}
  \usepackage[pdftex]{graphicx}

\else
  \usepackage[dvips]{graphicx}
\fi
\usepackage{subfigure}
\usepackage[cmex10]{amsmath}

\newcommand{\eqn}[1]{(#1)}

\newcommand{\fig}[1]{Fig.~#1}

\newcommand{\ie}{\mbox{\it i.e.}}

\newcommand{\spcend}{\ensuremath{\:}}

\newcommand{\vect}[1]{\ensuremath{\mbox{\boldmath ${#1}$}}}

\newcommand{\elmax}{\ensuremath{{L}}}

\newcommand{\nmeas}{\ensuremath{M}}

\newcommand{\sparmat}{\ensuremath{\Psi}}
\newcommand{\sensmat}{\ensuremath{\Phi}}

\begin{document}

\title{Implications for compressed sensing of a new\\sampling theorem
  on the sphere}

\author{
\IEEEauthorblockN{
Jason D. McEwen\IEEEauthorrefmark{1}, 
Gilles Puy\IEEEauthorrefmark{1}, 
Jean-Philippe Thiran\IEEEauthorrefmark{1},
Pierre Vandergheynst\IEEEauthorrefmark{1},
Dimitri Van De Ville\IEEEauthorrefmark{1}\IEEEauthorrefmark{2} and 
Yves Wiaux\IEEEauthorrefmark{1}\IEEEauthorrefmark{2}}
\IEEEauthorblockA{\IEEEauthorrefmark{1}
Ecole Polytechnique F{\'e}d{\'e}rale de Lausanne (EPFL), CH-1015
Lausanne, Switzerland}
\IEEEauthorblockA{\IEEEauthorrefmark{2}
University of Geneva (UniGE), CH-1211 Geneva,
Switzerland}
\thanks{This work is supported by CIBM of the Geneva
  and Lausanne Universities, EPFL, and the SNSF, Leenaards and Louis-Jeantet
  foundations.}
\vspace{-3mm}
}

\maketitle

Sampling theorems on the sphere state that all the information of a
continuous band-limited signal on the sphere may be contained in a
discrete set of samples.  For an equiangular sampling of the sphere,
the Driscoll \& Healy (DH) \cite{driscoll:1994} sampling theorem has
become the standard, requiring $\sim 4\elmax^2$ samples on the sphere
to represent exactly a signal band-limited in its spherical
harmonic decomposition at \elmax.  Recently, a new sampling theorem on an
equiangular grid has been developed by McEwen \& Wiaux (MW)
\cite{mcewen:fssht}, requiring only $\sim 2\elmax^2$ samples to
represent exactly a band-limited signal, thereby redefining
Nyquist rate sampling on the sphere.  No sampling theorem on the
sphere reaches the optimal number of samples suggested by the
$\elmax^2$ dimension of a band-limited signal in harmonic space
(although the MW sampling theorem comes closest to this bound).  A
reduction by a factor of two in the number of samples required to
represent a band-limited signal on the sphere between the DH and MW
sampling theorems has important implications for compressed sensing.

Compressed sensing on the sphere has been studied recently for signals
sparse in harmonic space \cite{rauhut:2011}, where a discrete grid on
the sphere is not required.  However, for signals sparse in the
spatial domain (or in its gradient) a discrete grid on the sphere is
essential.  A reduction in the number of samples of the grid required
to represent a band-limited signal improves both the dimensionality
and sparsity of the signal, which in turn affects the quality of
reconstruction.

We illustrate the impact of the number of samples of the DH and MW
sampling theorems with an inpainting problem, where measurements are made in
the spatial domain (as dictated by many applications).  A test signal
sparse in its gradient is constructed from a binary Earth map,
smoothed to give a signal band-limited at $\elmax=32$.  We first solve
the total variation (TV) inpainting problem directly on the sphere: \vspace*{-1.0mm}
\begin{equation}
\label{eqn:recon_spatial}
\vect{x}^\star =
\underset{\vect{x}}{\arg \min} \:
\| \vect{x} \|_{\rm TV} \:\: \mbox{such that} \:\:
\| \vect{y} - \sensmat \vect{x}\|_2 \leq \epsilon
\spcend ,
\end{equation}
where $\nmeas$ noisy measurements $\vect{y}$ of the signal $\vect{x}$
are made.  The measurement operator $\sensmat$ represents a random
masking of the signal.  The TV norm $\| \cdot
\|_{\rm TV}$ is defined to approximate the continuous TV norm on the
sphere and thus includes the quadrature weights of the adopted
sampling theorem, regularising the gradient computed on the sphere.
However, as discussed, the dimensionality of the signal $\vect{x}$ is
optimal in harmonic space.  Consequently, we reduce the
dimensionality of our problem by recovering the harmonic coefficients
$\vect{\hat{x}}$ directly:
\begin{equation}
\label{eqn:recon_harmonic}
\vect{\hat{x}}^\star =
\underset{\vect{\hat{x}}}{\arg \min} \:
\| \sparmat \vect{\hat{x}} \|_{\rm TV} \:\: \mbox{such that} \:\:
\| \vect{y} - \sensmat \sparmat \vect{\hat{x}}\|_2 \leq \epsilon
\spcend ,
\end{equation}
where $\sparmat$ represents the inverse spherical harmonic transform;
the signal on the sphere is recovered by $\vect{x}^\star = \sparmat
\vect{\hat{x}}^\star$.  For this problem the dimensionality of the
signal directly recovered $\vect{\hat{x}}$ is identical for
both sampling theorems, however sparsity in the spatial domain remains
superior (\ie\ fewer non-zero values) for the MW sampling theorem.

\begin{figure}
\centering
\includegraphics[width=75mm]{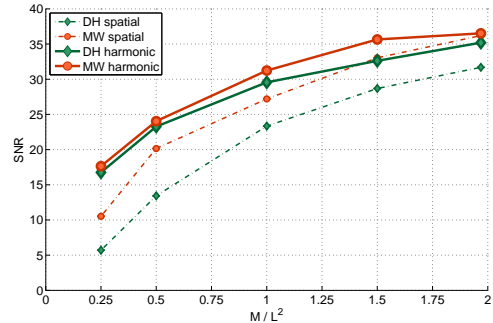}
\vspace{-3mm}
\caption{Reconstruction performance for the DH and MW sampling theorems}
\label{fig:snr_vs_m}
\end{figure}

Reconstruction performance is plotted in \fig{\ref{fig:snr_vs_m}} when
solving the inpainting problem in the spatial
\eqn{\ref{eqn:recon_spatial}} and harmonic
\eqn{\ref{eqn:recon_harmonic}} domains, for both sampling theorems
(averaged over ten simulations of random measurement operators and
independent and identically distributed Gaussian noise).  Strictly
speaking, compressed sensing corresponds to the range
$\nmeas/\elmax^2<1$ when considering the harmonic representation of
the signal.  Nevertheless, we extend our tests to $\nmeas/\elmax^2
\sim 2$, corresponding to the equivalent of Nyquist rate sampling on
the MW grid.  In all cases the superior performance of the MW sampling
theorem is evident.  In \fig{\ref{fig:spheres}} we show example
reconstructions, where the superior quality of the MW reconstruction
is again clear.

Although recovering the signal in the harmonic domain is more
effective, it is also computationally more demanding.  At present we
are thus limited to low band-limits.  To solve the convex optimisation
problem in the harmonic domain both the inverse spherical harmonic
transform and its adjoint operator are required.  A fast inverse
spherical harmonic transform exists \cite{mcewen:fssht}, from which a
fast adjoint operator follows directly.  The application of fast
inverse and adjoint operators is the focus of ongoing research and
will allow compressed sensing problems on the sphere to be tackled
effectively at much higher band-limits.

\newlength{\sphereplotwidth}
\setlength{\sphereplotwidth}{27mm}

\begin{figure}
\centering
\subfigure[Ground truth]{\includegraphics[clip=,viewport=2 2 438 223,width=\sphereplotwidth]{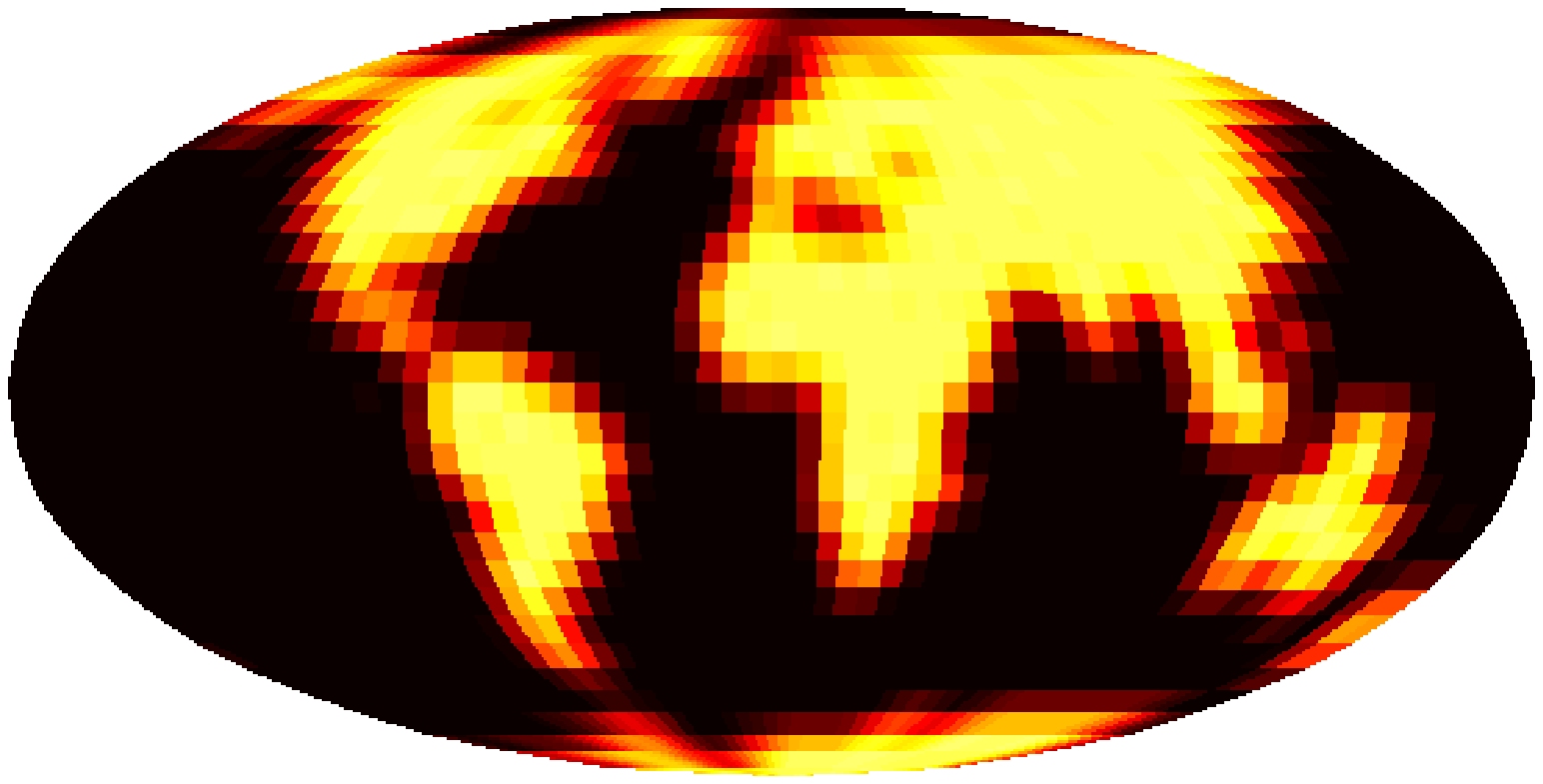}}
\subfigure[DH reconstruction]{\includegraphics[clip=,viewport=2 2 438 223,width=\sphereplotwidth]{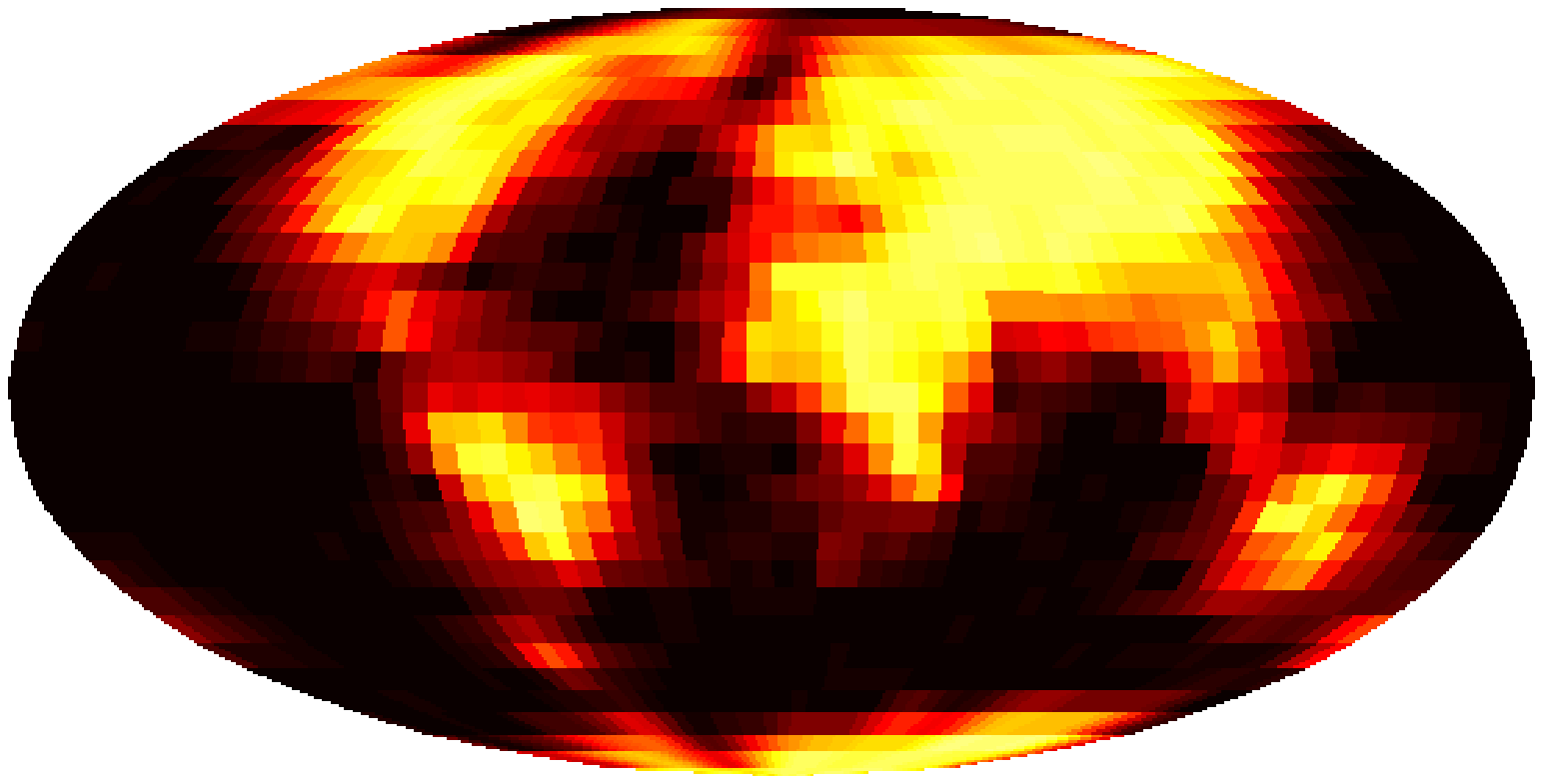}} 
\subfigure[MW reconstruction]{\includegraphics[clip=,viewport=2 2 438 223,width=\sphereplotwidth]{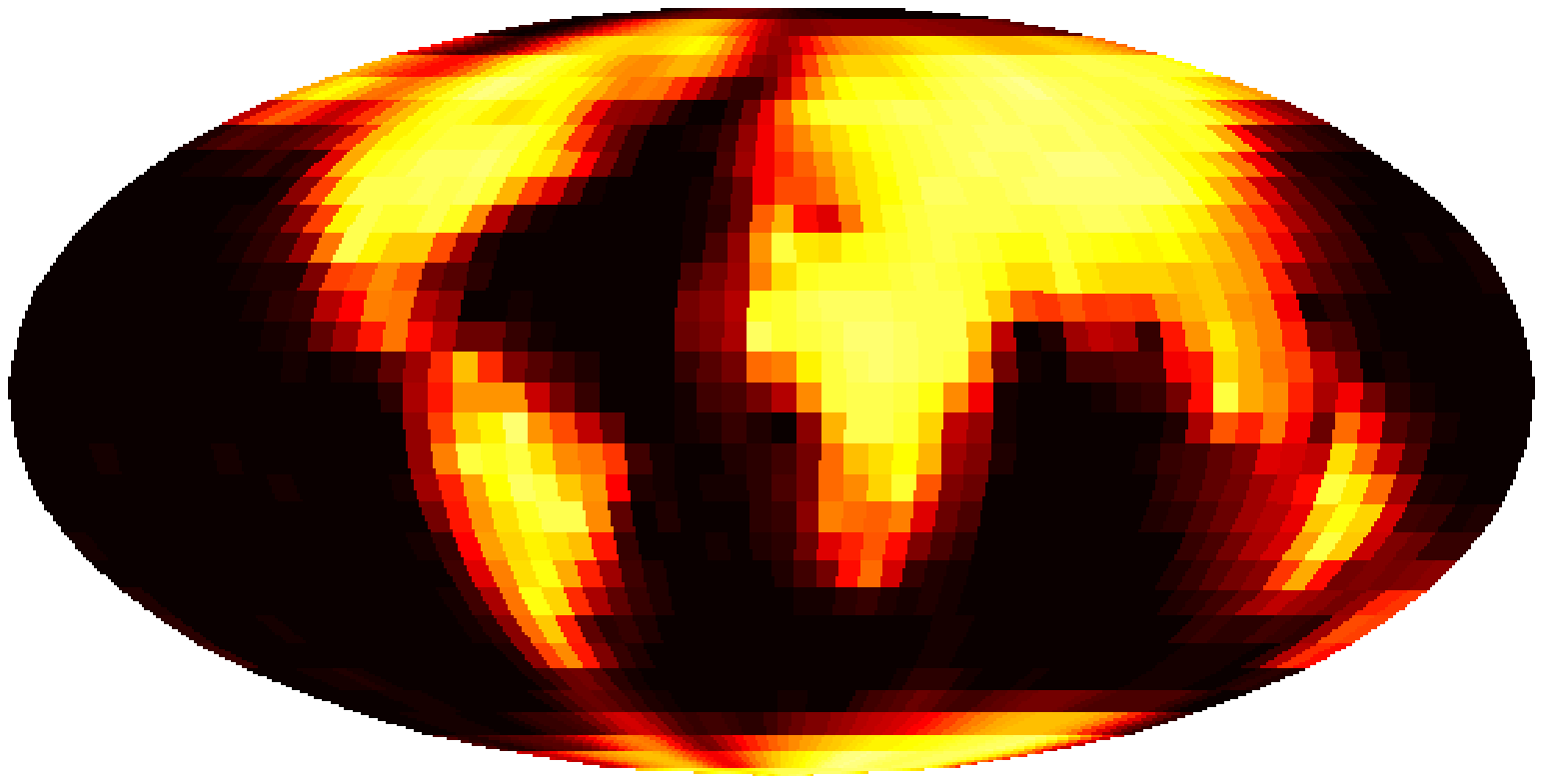}}
\caption{Reconstructed Earth topographic data for $\nmeas/\elmax^2 = 1/2$}
\label{fig:spheres}
\vspace{-4mm}
\end{figure}

\bibliographystyle{IEEEtran}
\bibliography{bib}

\end{document}